\documentclass[12pt]{iopart}
\usepackage[dvips]{graphicx}
\usepackage{psfrag}
\begin{document}

\title[FEM simulations of subwavelength apertures]{Accurate simulation of 
        light transmission through subwavelength apertures in metal films}

\author{D Lockau, L Zschiedrich and S Burger}

\address{Zuse Institute Berlin, Takustra{\ss}e 7, D-14195 Berlin, Germany}
\ead{burger@zib.de}
\begin{abstract}
We use the finite-element method for simulating light transmission 
through a 2D-periodic array of rectangular apertures in a film of 
highly conductive material. 
We report results with a relative error of the transmissivity lower than 
0.01\%.
This is an improvement of about one order of magnitude compared to 
previously reported results. 
Further, the influence of corner and edge 
roundings on light transmission through the subwavelength apertures is 
investigated. 
\end{abstract}

\normalsize
\section{Introduction}
Experiments investigating light transmission through subwavelength 
apertures in metallic films have revealed surprising 
phenomena like enhanced transmission and beaming effects~\cite{Ebbesen1998a,Lezec2002a}. 
The mechanisms involved in the transmission spectra can  
be explained using 2D models and approximative methods ~\cite{MartinMoreno2003a}. 
For a fully quantitative 
understanding and from the point of view of a nano-optical design of these structures, 
accurate numerical simulations 
of Maxwell's equations for cylindrically symmetric problems as well as for problems 
with full 3D geometry are desired~\cite{Popov2005a,Granet2006a}.

A variety of different methods exist for rigorous simulations of Maxwell's equations. 
Among these are the finite-element method (FEM), the finite-difference time-domain method (FDTD), 
the Fourier-modal method (FMM, RCWA), the boundary-element method (BEM), and others. 
However, especially in cases of large 3D computational domains and of highly conductive 
materials (like silver or gold) the field distributions of interest are so complicated that 
it becomes challenging for all numerical methods to accurately approximate them. 
Recently, Granet and Li~\cite{Granet2006a} have presented results on the convergence for 
the simulation of light transmission through periodically perforated thin silver films. 
They use a dedicated Fourier-modal method which allows for spatial adaptivity and for the use of 
symmetries of the problem. This allows them to obtain numerical results which are converged to  
a relative error of the order of 0.1\%~\cite{Granet2006a_note1}.
In their paper, Ganet and Li suspect that all other published numerical results obtained so far 
for the same type of problems are not fully converged. 
I.e., due to the lack of efficient methods, many published results regarding  
numerical simulations of Maxwell's equations for nanooptical problems 
are at improper accuracy levels. 

In this contribution we revisit the same problem using the finite-element method. 
We show that our method yields the same numerical values within  numerical uncertainty, and 
we present convergence results showing an obtained relative error which is about one order of magnitude 
lower than the results of Granet and Li. 
Further we investigate the influence of corner roundings on the transmissivity of the sample. 


\section{Problem statement}
\label{chapter_setup}
\begin{figure}[t]
  \begin{center}
    \includegraphics[width=.8\textwidth]{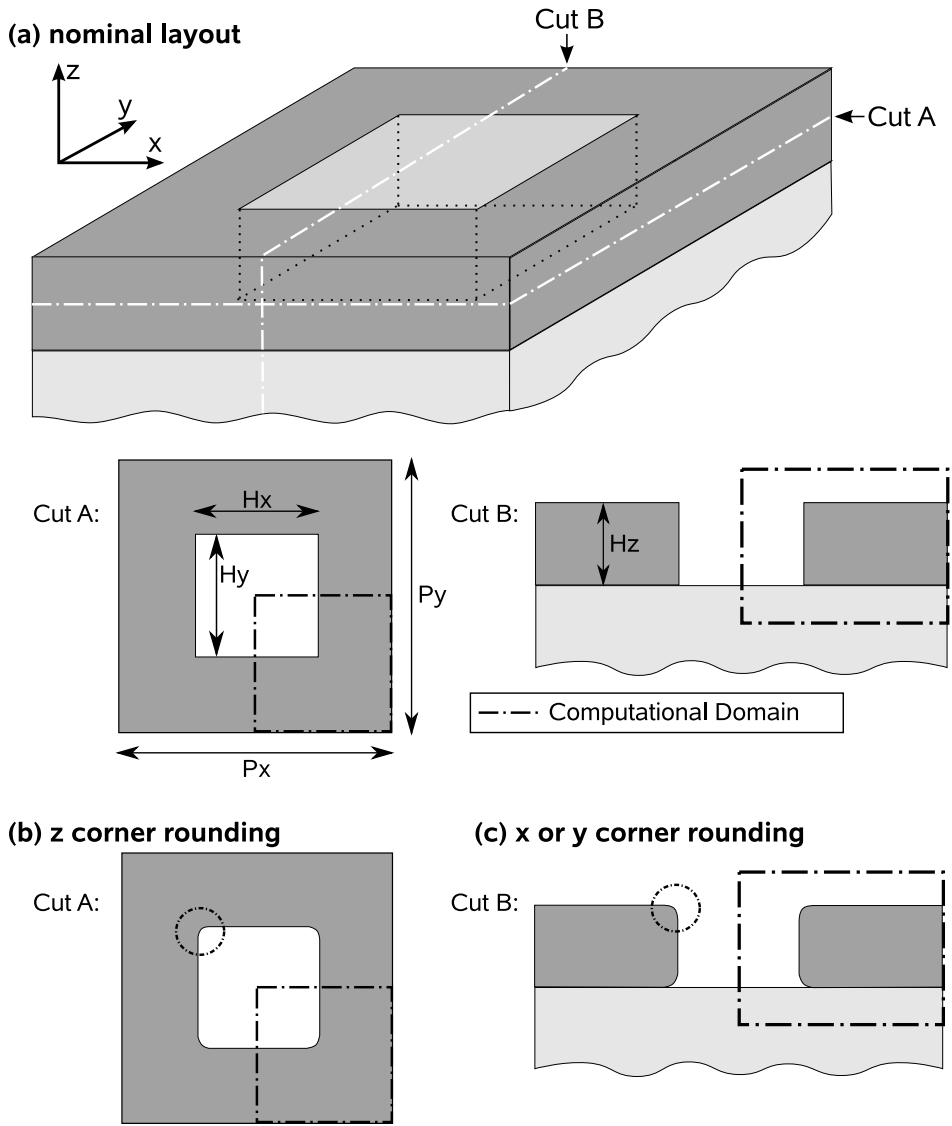}
  \end{center}
  \caption{Unit cell of the periodic grating layout used in the numerical 
    experiments. (a) Layout without corner rounding. \textit{Computational Domain:}
    The twofold mirror symmetry of the unit cell allows a 
    reduction of the domain on which the solution is computed. (b)
    Corner rounding on all $z$ edges. (c) Corner rounding on all $z$
    as well as all $x$ and $y$ edges. 
    The substrate is displayed in light grey, the silver layer is displayed in dark grey, and 
    vacuum is displayed in white.}
  \label{fig:setup}
\end{figure}

\begin{figure}[t]
  \centering
  \includegraphics[width=0.6\textwidth]{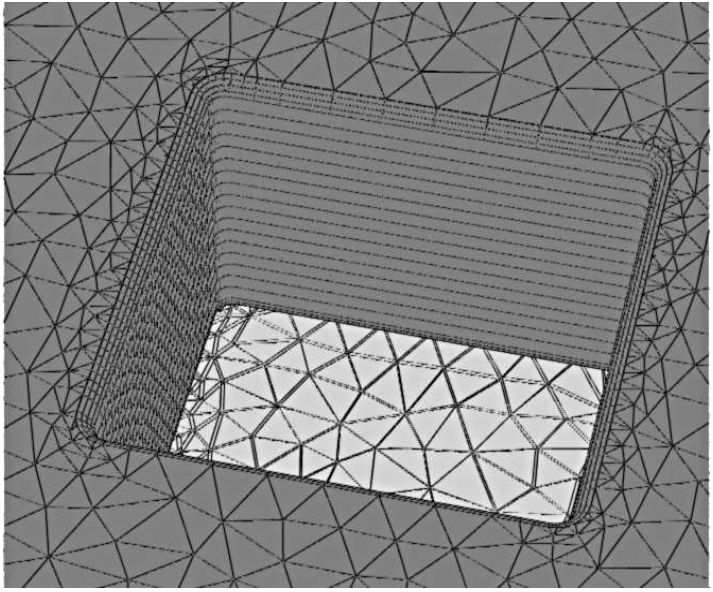}
  \caption{Detail of a graphical representation of the discretized silver layer with a 
   square shaped aperture with corner roundings of $10\,\textrm{nm}$ radius on all edges
   (compare Figure~\ref{fig:setup} c).}
  \label{fig:mesh}
\end{figure}

The exemplary geometric layout we study in this paper has
already been thoroughly examined by Granet and Li \cite{Granet2006a}.
Despite its geometric simplicity it turns out to be a challenging case
in the achievement of sufficient numerical convergence of the solution.
The layout consists of a doubly periodic, infinite array of square-shaped 
apertures in a highly
conductive metal layer on a glass substrate, c.f. Fig.~\ref{fig:setup}. 
Monochromatic light is incident in $-z$-direction,
linearly polarized in $x$-direction, and with 
a vacuum wavelength of $\lambda_0 = 1.3875\,\mu$m.
A unit cell of the periodic layout and 
the computational domain are shown in Fig.~\ref{fig:setup}, (a).
The substrate consists of glass with a relative permittivity
$\varepsilon_{\textrm{\small Glass}} = 1.5^2$. 
The silver grating 
has a relative permittivity of $\varepsilon_{\textrm{\small Silver}} = (0.1+8.94i)^2$
at the wavelength of interest. 
The half space above the grating as well as the apertures have the
material properties of vacuum, $\varepsilon_{\textrm{\small vac}} = 1.0$.
The thickness of the metal layer is $\textrm{Hz} = 200\,\textrm{nm}$, the lateral
dimensions of the square apertures are $\textrm{Hx}=\textrm{Hy} =
250\,\textrm{nm}$, and the periodicity lengths of 
the layout are $\textrm{Px}=\textrm{Py}=900\,\textrm{nm}$.
This combination of geometry, materials and incident light properties results in 
enhanced light transmission through the apertures (see Section~\ref{sec:enhanced_transmission}).

We also investigate the influence of corner roundings 
on the transmissivity through the sample. 
Fig.~\ref{fig:setup}, (b) schematically shows how corner rounding is applied 
to edges parallel to the  $z$-axis of the coordinate system, and 
 Fig.~\ref{fig:setup}, (c) shows additional corner roundings applied 
to edges parallel to the  $x-$ and $y$-axes (compare also Fig.~\ref{fig:mesh}).


\section{Finite-element method}
We use the finite-element method to find rigorous solutions to the linear, time-harmonic 
Maxwell's equations. In a nutshell, the method consists of the following steps:
\begin{itemize}
\item
The geometry of the computational domain is discretized with simple
geometrical patches.
2D layouts are discretized using, e.g., triangular patches. For a 3D layout
tetrahedral, prismatoidal or other patches are used.
In the simulation results presented here we have used prismatoidal patches.
Transparent boundary conditions (BC) are applied in the $\pm z$-directions, mirror-symmetric
boundary conditions are applied in the lateral directions. 
Alternatively, the method allows for Bloch-periodic or transparent BC
also in the lateral directions. 
Transparent BC are realized with a perfectly matched layer (PML) method~\cite{Zschiedrich2008al}. 
\item
The function spaces in the integral representation of Maxwell's equations
are discretized using Nedelec's edge elements,
which are vectorial functions of polynomial order,
defined on the simple geometrical patches~\cite{Monk2003a}.
In the current implementation, our FEM solver uses polynomials of
first to ninth order.
FEM can be explained as expanding the field
corresponding to the exact solution of Maxwell's equations in the
basis given by these elements.
\item
The expansion leads to a large sparse matrix equation (algebraic problem).
To solve the algebraic problem on a standard workstation
linear algebra decomposition techniques (e.g., sparse LU-factorization)
are used.
For the case when the available memory (RAM) of the computer becomes the limiting factor 
for large computational domains or high accuracy
demands, we have also implemented rigorous domain decomposition methods~\cite{Zschiedrich2008al}.
These allow to handle problems with larger numbers of unknowns.
However, all results presented in this paper are obtained by sparse LU-factorization. 
\end{itemize}

For details on our FEM implementation, e.g.,
the choice of Bloch-periodic functional spaces,
and the adaptive FEM discretization 
we refer to previous works~\cite{Pomplun2007pssb}.
Our solver has been compared and benchmarked with RCWA and FDTD-based Maxwell solvers
for 2D~\cite{Burger2005bacus} and
3D~\cite{Burger2006c,Burger2008bacus} computational domain problems. 
It has been successfully applied to various problem classes, e.g., to simulate properties of 
metamaterials, photonic crystal fibers, plasmonic structures, 
nanoresonators, and integrated optical devices. 
We note that our finite-element solver is also part of a commercial program
package~\cite{cite_jcmwave_jcmsuite}.


\section{Numerical investigations}
\subsection{Light transmission through apertures with sharp edges}
\label{sec:num_ex_sharp}

\begin{figure}[htb]
\psfrag{T}{\sffamily T}
\psfrag{N}{\sffamily N [$10^6$]}
\psfrag{0.09}{\sffamily 0.09}
\psfrag{0.08}{\sffamily 0.08}
\psfrag{0.07}{\sffamily 0.07}
\psfrag{0}{\sffamily 0}
\psfrag{2}{\sffamily 2}
\psfrag{4}{\sffamily 4}
\psfrag{6}{\sffamily 6}
\centering
\includegraphics[width=0.5\textwidth]{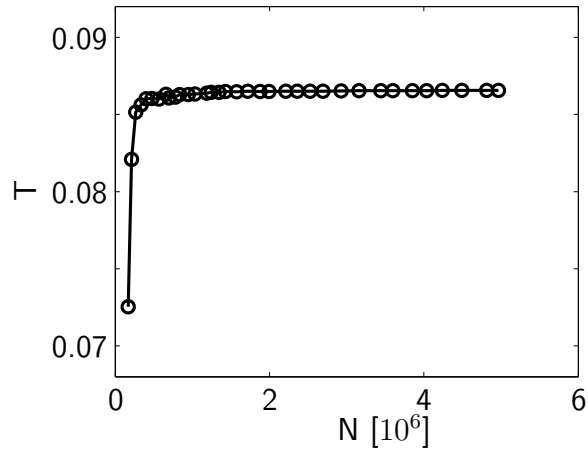}
\caption{Convergence of the simulated transmittivity 
$T$ in dependence on the number 
of unknowns $N$ of the FEM problem.}
\label{figure_conv_p2}
\end{figure}

We have performed FEM computations on meshes with increasing refinement in order to 
investigate the convergence behavior of our method for the simulation example described in 
Chapter~\ref{chapter_setup} and Figure~\ref{fig:setup} (a). 
The transmissivity through 
the periodic array of apertures, $T$, is computed from the finite-element near field solutions.
The number of degrees of 
freedom of the algebraic problem, $N$, increases with increasing refinement of the prismatoidal 
mesh discretizing the geometry. 
Figure~\ref{figure_conv_p2} shows the dependence of the absolute value of transmissivity through 
the periodic array of apertures, $T$, on the number of unknowns, $N$.
As expected, we obtain convergence of the solution with increasing $N$.
The results presented here are obtained using finite elements of polynomial degree $p=2$. 
The absolute value of transmissivity converges to a value of about 0.0866, 
see Tab.~\ref{tab:cr_summary}. 
Granet and Li computed the same number using RCWA, within the error bounds of the 
numerical solutions~\cite{Granet2006a_note1}. 

\begin{figure}[htb]
\centering
\includegraphics[width=0.5\textwidth]{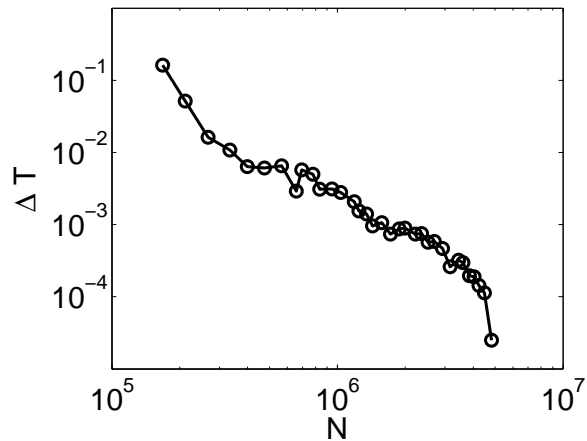} %
\caption{Relative error of the transmittivity, $\Delta T$,  
in dependence on the number of unknowns  $N$ of the FEM problem.}
\label{figure_conv_rel_p2}
\end{figure}

Figure~\ref{figure_conv_rel_p2} shows the convergence of the relative error of 
 transmissivity, $\Delta T = |T(N)-T_{qe}|/|T_{qe}|$, corresponding to the results shown in 
Figure~\ref{figure_conv_p2}. 
Here $T_{qe}$ is the transmissivity computed from the quasi-exact solution, i.e., 
from a solution obtained with our FEM solver 
on a finer mesh than the meshes of the solutions corresponding to $T(N)$. 
The numerical value of the transmissivity computed from the quasi-exact solution is 
quoted in Table~\ref{tab:cr_summary} ($R_z=0$, $R_{xy}=0$).
The fine mesh of the quasi-exact solution results in  
a number of unknowns of the FEM problem of $N=4,967,892$. 
Indeed, it would be desirable to compare $T(N)$ to an analytical solution. 
However, for problems where an analytical solution is not available, the quasi-exact solution 
is used as a makeshift. 
From Fig.~\ref{figure_conv_rel_p2} it can be seen that solutions with a relative accuracy of 
transmissivity of about $1\%$ 
can be obtained with a computational effort of a few $10^5$ unknowns. 
Relative accuracy of about $0.1\%$ is reached with about one million of unknowns, and an 
accuracy of 
about $10^{-4}$ can be obtained with a computational effort of some millions of unknowns. 
Computation times range from below a minute on the 1\%-accuracy side to about one hour for 
the best converged solutions. Memory (RAM) consumption for the displayed results 
ranges between below 1GB and about 100GB, again depending on accuracy. 
We note that the computations have been performed on a standard workstation with extended RAM 
which recently has become affordable due to the price decline of modern computers. 
However, when RAM consumption becomes the limiting factor for high precision computations, 
it is also possible to compute solutions on fine grids using domain-decomposition 
algorithms~\cite{Zschiedrich2008al} instead 
of the direct LU factorization of the full problem matrix. Therefore, 
all presented results can also 
be obtained on desktop computers with limited RAM. 

We further note that keys for achieving highly converged results using FEM are, firstly, 
the adaptive discretization of the geometry where the mesh is finer close to corners and edges 
of the geometry, and, secondly, a careful implementation of transparent boundary conditions.

\subsection{Light transmission through apertures with corner rounding}
\label{sec:zcr}

The good convergence properties of FEM allow us to investigate the 
influence of small changes in the geometry on the transmissivity. 
In this section we investigate the convergence properties of the method when corner rounding is 
introduced. 
Further we observe that the transmissivity can be greatly influenced by corner roundings 
of the structure. 

\begin{figure}[t]
  \centering
  \psfrag{T}{\sffamily T}
  \psfrag{N}{\sffamily N [$10^6$]}
  \psfrag{0}{\sffamily 0}
  \psfrag{1}{\sffamily 1}
  \psfrag{2}{\sffamily 2}
  \psfrag{0.04}{\sffamily 0.04}
  \psfrag{0.05}{\sffamily 0.05}
  \psfrag{0.06}{\sffamily 0.06}
  \psfrag{0.07}{\sffamily 0.07}
  \psfrag{0.08}{\sffamily 0.08}
  \psfrag{0.09}{\sffamily 0.09}
  \psfrag{c-radius--2-nm}{\sffamily corner radius: $2   \textrm{nm}$}
  \psfrag{c-radius--10-nm}{\sffamily corner radius: $10 \textrm{nm}$}
  \psfrag{c-radius--20-nm}{\sffamily corner radius: $20 \textrm{nm}$}
  \psfrag{no-c-rounding}{\sffamily no corner rounding}
  \includegraphics[width=0.6\linewidth]{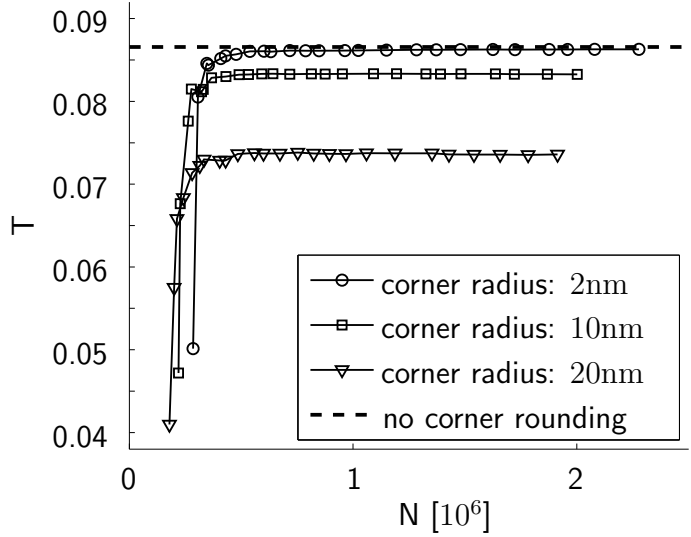}
  \caption{Transmissivity over  number of unknowns for different roundings of $z$ edges. The dashed horizontal line shows the value to which the layout without corner rounding has converged.}
  \label{fig:zcr_convergence}
\end{figure}

We first introduce a rounding of the $z$ edges of the apertures,
cf. Fig. \ref{fig:setup}(b). 
The rounding is approximated by 13 polynomial segments per quarter circle. 
As in the previous section we compute FEM solutions for different mesh 
qualities for a convergence analysis. 
To investigate the influence of corner rounding we choose different rounding radii between 
0.1\,nm and 20\,nm. 
Figure \ref{fig:zcr_convergence} provides numerical results for three different corner radii. 
Displayed is again the transmissivity in dependence on the number of unknowns. 
Again we observe very good convergence. 
Table \ref{tab:cr_summary} gives a summary of the transmissivity values for a finer sampling of 
corner radii. 
All indicated numerical errors are obtained from a convergence analysis as described above. 
We observe that the transmissivity value changes only by 
$<1\%$ for a corner rounding of $3\,\textrm{nm}$ and below and it changes by up to 15\% for 
a corner rounding of 20\,nm.

\begin{table}[t]
  \centering
  \begin{tabular}{|l|l|l|}
    \hline\hline
    $R_z$ [nm] & $R_{xy}$ [nm] & T\\
    \hline
    0.0& 0.0& $8.6564\times 10^{-2} \pm 8\times 10^{-6}$  \\
    \hline
    0.1& 0.0& $8.650\times 10^{-2}  \pm 2\times 10^{-5} $ \\
    0.5& 0.0& $8.647\times 10^{-2}  \pm 2\times 10^{-5} $ \\
    1.0  & 0.0& $8.644\times 10^{-2}  \pm 3\times 10^{-5} $ \\
    2.0  & 0.0& $8.6277\times 10^{-2} \pm 3\times 10^{-6} $ \\
    3.0  & 0.0& $8.6099\times 10^{-2} \pm 8\times 10^{-6} $ \\
    5.0  & 0.0& $8.5584\times 10^{-2} \pm 7\times 10^{-6} $ \\
    7.0  & 0.0& $8.4842\times 10^{-2} \pm 3\times 10^{-6} $ \\
    10.0 & 0.0& $8.326\times 10^{-2}  \pm 3\times 10^{-5} $ \\
    20.0 & 0.0& $7.360\times 10^{-2}  \pm 6\times 10^{-5} $ \\
    \hline
    2.0 & 2.0  &  $7.31\times 10^{-2}$ \\
    10.0 & 10.0  & $3.24\times 10^{-2}$ \\
    \hline\hline
  \end{tabular}
  \caption{Transmissivity values for layouts with different corner
    radii on the vertical edges ($R_z$) and on the horizontal edges ($R_{xy}$).}
  \label{tab:cr_summary}
\end{table}

In a next set of numerical experiments 
we introduce corner rounding on all edges, 
{\it cf.,} Figs.~\ref{fig:setup}(c) and~\ref{fig:mesh}. 
We choose two corner radii, $2\,\textrm{nm}$ and $10\,\textrm{nm}$ for comparison.
The transmissivity values in those cases are
also included in Table~\ref{tab:cr_summary}. 
A larger impact of the rounding on the horizontal edges on transmissivity is observed:
For the $10\,\textrm{nm}$ corner rounding the transmissivity value drops 
by a factor of less than 50\% of the aperture with sharp edges and of the apertures with 
$R_z\leq 10$\,nm.

These results show that for quantitative studies of optical properties of metallic nanostructures 
corner roundings in general have to be taken into account, even for relatively small rounding radii.
Such roundings are always present in the experimental setups, and as shown they can have a 
critical impact on the performance of nanooptical devices.


\subsection{Enhanced light transmission}
\label{sec:enhanced_transmission}

\begin{figure}[htb]
\psfrag{Transmission}{\sffamily T}
\psfrag{Pitch [nm]}{\sffamily Px [nm]}
\psfrag{0.1}{\sffamily 0.1}
\psfrag{0.08}{\sffamily 0.08}
\psfrag{0.06}{\sffamily 0.06}
\psfrag{0.04}{\sffamily 0.04}
\psfrag{0.02}{\sffamily 0.02}
\psfrag{0}{\sffamily 0}
\psfrag{800}{\sffamily 800}
\psfrag{850}{\sffamily 850}
\psfrag{900}{\sffamily 900}
\psfrag{950}{\sffamily 950}
\psfrag{1000}{\sffamily 1000}
\psfrag{895}{\sffamily 895}
\psfrag{905}{\sffamily 905}
\centering
a) \includegraphics[width=0.45\textwidth]{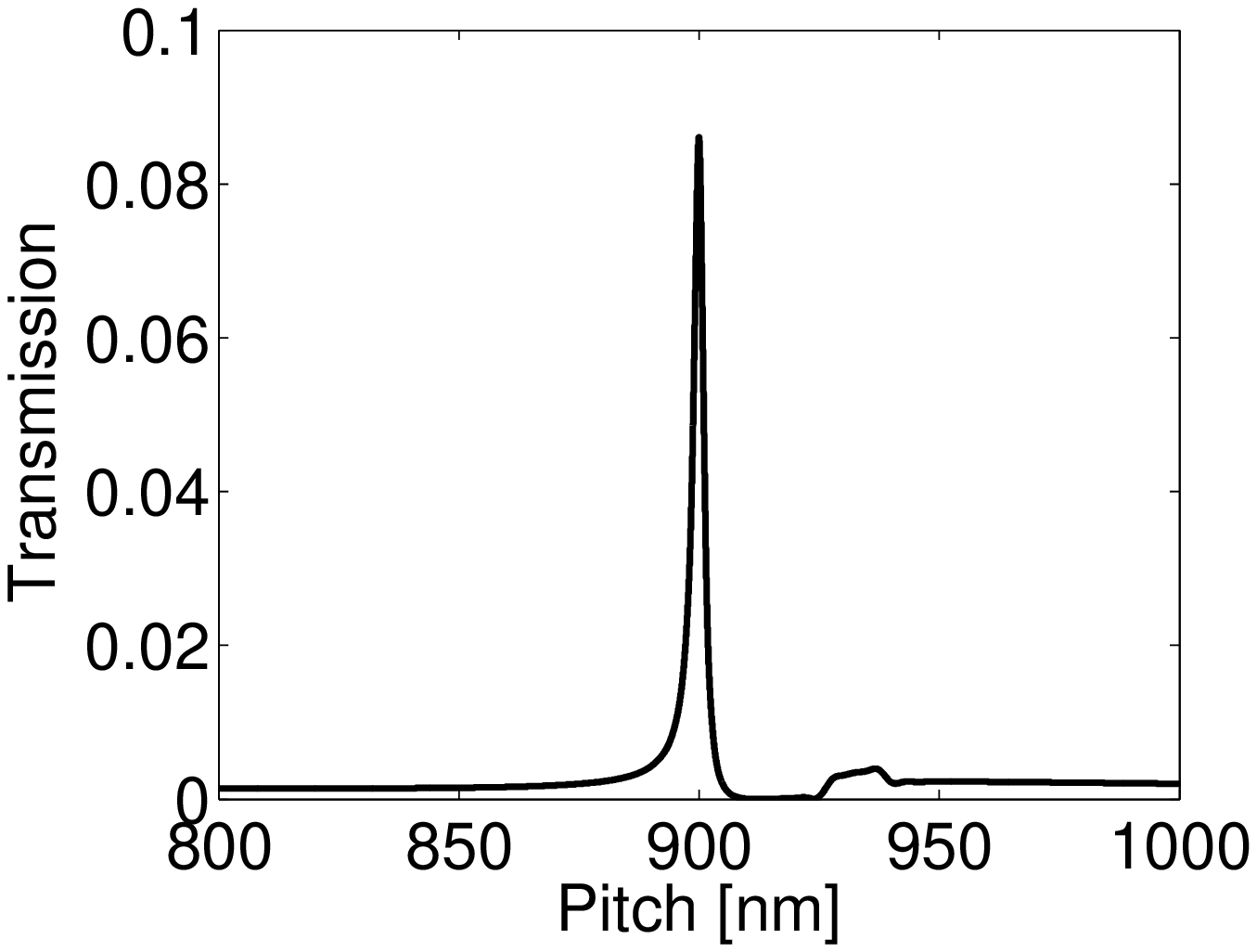}\\
b) \includegraphics[width=0.45\textwidth]{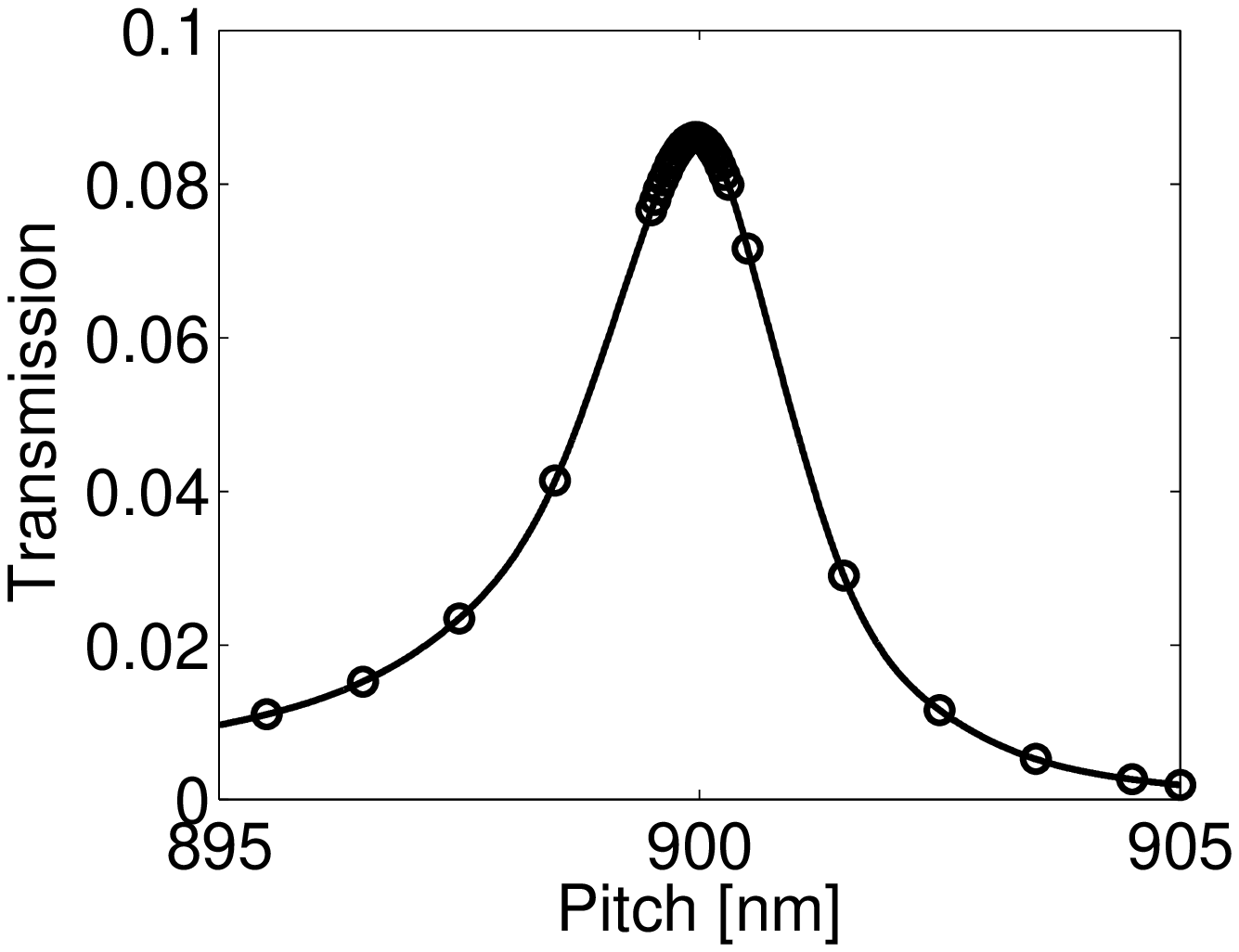}
\caption{
Transmission $T$ in dependence on periodicity length 
$\mathrm{P}$.
(a) Enhanced transmission appears as a very sharp peak in the scanned parameter range.
(b) Detail with the peak shape
(Circles: simulation results; solid line: spline fit).}
\label{figure_enhanced}
\end{figure}

In order to understand processes involved in light transmission through the investigated array of 
holes we have performed the following numerical experiment: 
We varied the periodicity length $\textrm{P}$ ($\textrm{P} = \textrm{Px} = \textrm{Py}$)
of the array and recorded the simulated transmissivity. 
For this scan we have used second order finite elements and discretizations with about 
$5\times 10^5$ unknowns which corresponds to a relative error of $\Delta T < 1\%$  
({\it cf.}, Section~\ref{sec:num_ex_sharp}).
Figure~\ref{figure_enhanced} shows the transmission through the array in the range $\textrm{P} = 800\dots 1000$\,nm. 
A sharp transmission peak is observed
with a full width at half maximum of only about 2.5\,nm, centered around a periodicity length
of $\textrm{P} = 900$\,nm.

\begin{figure}[htb]
\centering
\includegraphics[width=0.95\textwidth]{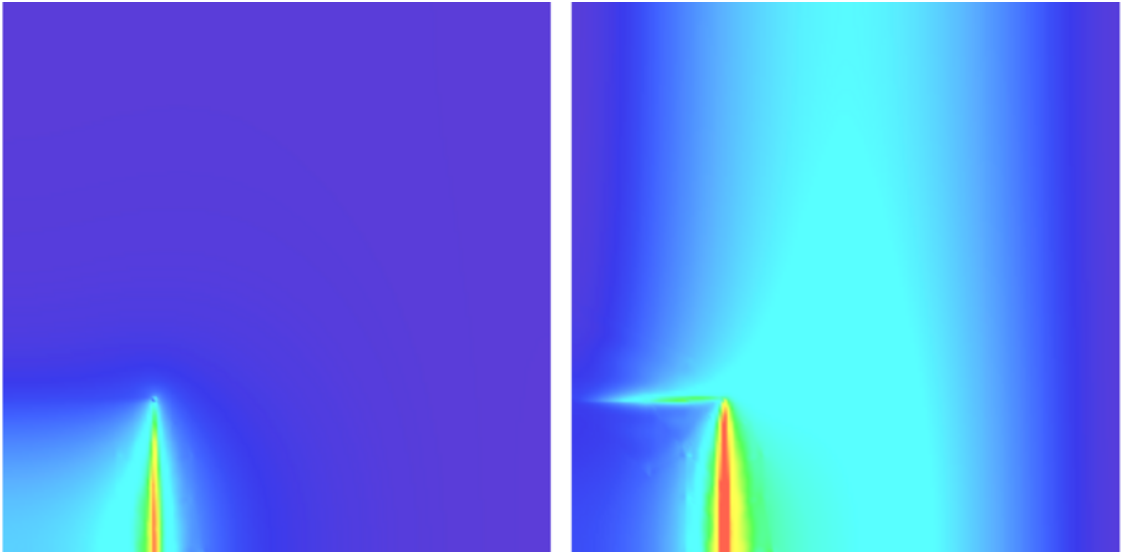}
\caption{Electric near field amplitudes in $x-y-$sections at different heights $z$ (left: 
1\,nm above the metal film, 
right: 1\,nm below the metal film).
Pseudocolor plots of $|E|$ with linear scales blue to red:
0 \dots $1.5\,\mathrm{[V/m]}$ (left), resp.\
0 \dots $15.0\,\mathrm{[V/m]}$ (right).
(\it See original publication for images with higher resolution.}
\label{figure_nf_xy}
\end{figure}

We have exported sections through the 3D field distributions in order to visualize the 
electric field distribution at peak transmission ($\textrm{P} = 900$\,nm). 
Figure~\ref{figure_nf_xy} shows sections parallel to the metal film, 1\,nm above the film
 and 1\,nm below the film. 
Plotted is the amplitude of the electric field. The sections are taken through one quarter of 
the unit cell of the periodic array. 
The scales of the pseudocolor plots range between 0 to 1.5\,V/m (above the metal film)
and 0 to 15\,V/m (below the metal film).
Please notice that the exciting light field is incident from the top. 
At first sight it is surprising that the near field intensity is orders of 
magnitude larger at the bottom 
side of the array, i.e., on the {\it `shaded'} side of the structure. 
This can explained by the fact that for the  resonant structure 
counterpropagating surface plasmons are excited at the 
metal-substrate interface.\cite{Popov2000a} 
The surface plasmons propagate in $\pm x$-direction (as can be seen, e.g., from the standing wave pattern in 
Figure~\ref{figure_nf_xy}, right).
The explanation of enhanced transmission through the excitation of surface plasmons is also 
in accordance with the narrow width and the position of 
the peak in the transmission spectrum ({\it cf.}, Fig.~\ref{figure_enhanced}): 
We have checked that surface plasmons on an unstructured material stack are excited in a Kretschmann configuration 
for the investigated material parameters and exciting
wavelength.\cite{Novotny2006a} 
The surface plasmons do not contribute directly to transmission in forward-direction, 
but they are scattered at the 
holes in the metal film which causes large energy flux into the substrate. 

Also the larger impact of rounding of the horizontal edges compared to rounding of the 
vertical edges (as investigated in Section~\ref{sec:zcr}) is in accordance with the explanation 
of enhanced transmission caused by a plasmonic surface state of the structure:
The scattering cross-section for scattering of surface plasmons at holes in the metal film 
depends critically on the geometry of the holes, 
especially at the metal-glass interface where the surface plasmon amplitude is at its maximum.

\section{Conclusion}
We have presented a finite-element method for simulating light transmission through periodic arrays of 
apertures in a highly conducting metal films. 
Well converged simulation results have been achieved. 
Despite the demonstrated and theoretically expected advantage in  
convergence properties, FEM is a 
very general and flexible method which also allows to investigate more realistic 
problem settings. 
Corner roundings, sidewall angles or other small geometrical features can be a roadblock for 
accurate simulations with other methods relying, e.g., on more regular grids.
In contrast, such geometrical features 
can very well be resolved using finite-element methods on unstructured grids. 
We have demonstrated this advantage by investigating the influence of corner and edge 
roundings on light transmission through subwavelength apertures. 

\section*{Acknowledgements}
We acknowledge fruitful discussions with B.\,H.~Kleemann and thank him for pointing out 
the work of Granet and Li~\cite{Granet2006a} to us.  

\section*{References}
\bibliography{phcbibli,group,lithography}

\begin{thebibliography}{10}
\bibitem{Granet2006a_note1}
From Figure~3 of Reference [5] we estimate a simulated transmissivity of
  $0.0865 \pm 0.0001$ which corresponds to a relative error of about 0.1\%.
\end{thebibliography}
\bibliographystyle{spiebib}

\end{document}